# Web 3.0 Adoption Behavior: PLS-SEM and Sentiment Analysis

Sheikh M. Hizam[1], Waqas Ahmed[*,1], Habiba Akter[1], Ilham Sentosa[1] and Mohamad N. Masrek[2]

[1] *UniKL Business School (UBIS), Universiti Kuala Lumpur, Kuala Lumpur, 50300, Malaysia.*
[2] *Faculty of Information Management, Universiti Teknologi MARA (UiTM), Selangor, 40450, Malaysia*

**Abstract**
Web 3.0 is considered as future of Internet where decentralization, user personalization and privacy protection would be the main aspects of Internet. Aim of this research work is to elucidate the adoption behavior of Web 3.0 through a multi-analytical approach based on Partial Least Squares Structural Equation Modelling (PLS-SEM) and Twitter sentiment analysis. A theoretical framework centered on Performance Expectancy (PE), Electronic Word-of-Mouth (eWOM) and Digital Dexterity (DD), was hypothesized towards Behavioral Intention (INT) of the Web 3.0 adoption. Surveyed data were collected through online questionnaires and 167 responses were analyzed through PLS-SEM. While 3,989 tweets of "Web3" were analyzed by VADER sentiment analysis tool in RapidMiner. PLS-SEM results showed that DD and eWOM had significant impact while PE had no effect on INT. Moreover, these results were also validated by PLS-Predict method. While sentiment analysis explored that 56% tweets on Web 3.0 were positive in sense and 7% depicted negative sentiment while remaining were neutral. Such inferences are novel in nature and an innovative addition to web informatics and could support the stakeholders towards web technology integration.

**Keywords**
Web 3.0, Adoption Behavior, PLS-SEM, Sentiment Analysis

## 1. Introduction

Internet has become essential part of daily life and its progressive nature is renovating and upgrading the ways of society, business, academia, and governance function. Web technologies for social connectedness are overwhelmingly accepted by each group of society where matters of interest and concerns are being communicated and compensated by emotions and monetary means e.g., Facebook, Twitter, Instagram, Airbnb, etc. However, three decades ago, this web infrastructure was not in such advanced form. The Internet was initiated as read-only format i.e., Web 1.0, where websites were used to merely display certain information and Internet users had no facilities to write on Internet platform. With the integration of innovative mindset and development in technological process had led Web 1.0 to the Web 2.0, where read and write functions were facilitated for users in shape of blogs, posts, comments, feeds, and tweets etc. For instance, current electronic social integration mechanism that has enabled the Internet users to read, write, share, and then impact the society, businesses, and governments by using Internet and social media. Web 2.0 was mainly based on mobile technology and social media as invention and integration of smartphones and Facebook, Orkut, Twitter etc. were orchestrated around same time. Moreover, cloud technology also transfigured Web 1.0 to turn into Web 2.0.

It's been more than a decade since Web 2.0 is transforming the lives across the globe specifically in global north while global south is still entrenching the digital mechanism to receive all the privileges of Web 2.0 i.e., computer/laptop/smartphone facilities, Internet connectivity, access to information and social media, learning opportunities etc. However, Web 2.0 has certain limitation for Internet users such as centralized control of information dissemination and access, poor security and safety system, lack of personalization and privacy hacks. To overcome the limitations of current web technologies, the concept of Web 3.0 is considered as solution platform.

As Web 3.0, hereinafter referred to as Web3, concept is reiterated mostly over the Internet and social media by entrepreneurs and tech-savvy professionals [1], [2] and few studies have also strived to define and explain this term [3]–[5]. Skimming the description of Web3 through available literature, it can be described as the advanced form of Web 2.0 which is decentralized in nature by blockchain technologies, personalized by the interactions of users and supported the individuality and privacy of its users. Moreover, it was also reiterated that control over personal and professional data of each user will be handled by him/herself through blockchain tokens where governments and tech-giants will no longer have privileges to interfere, for instance, the usability of Non-Fungible Tokens (NFTs) and smart contracts. Similarly, the Decentralized Finance (DeFi) concept has also floated around Web3 through cryptocurrencies based on secure distributed ledgers.

The humming of Web3 is also integrated after "Metaverse" initiative over Internet which is novel concept introduced by Facebook towards future network of 3D virtual world centered on social connection [6], [7]. However, as this emerging concept has been buzzing around the Internet since many years, various web companies (i.e., YouTube, Facebook, Twitter, Instagram etc.) have also started the implementation of personalizing the Internet for users through Artificial Intelligence through collecting their interaction data, visit pathways (website cookies), and their preferences towards society, food, shopping, education, business, skills etc. This may be the initial phase of instigation of Web3. However, being the future of Internet, there is a paucity of research work that could describe the interaction and adoption behavior of Web3 to understand the behavioral pattern and usability mechanism of end-users.

Understanding the behavior and acceptance sense of Web3 would yield the novel contribution to the knowledge and would pave the way for broad and concrete research opportunities in human-technology interaction field. Largely users share their verbatim views over social media platform such as Twitter, which can be positive or negative in nature and by analyzing the sentiments in their tweets could also reveal the overall impression towards respective phenomenon [8]. Twitter sentiment analysis is a useful method to understand the polarity of sentiment in users tweets towards any topic [9]. Similarly, technology adoption behavior mainly focuses over the projected productivity of respective digital service referred to as Performance Expectancy [10], [11], the cognitive stimuli to use technology with competence and awareness of digital tools usability and benefits that conveys to the expression of Digital Dexterity [11], [12], and society's views of approval and disapproval towards certain digital mechanism in shape of Electronic Word-of-Mouth [13], [14]. By contemplating these factors into a theoretical framework and hypothesizing their relationship towards Web3 usability would support to conclude the adoption behavior of Web3.

By summarizing the rationale of Web3 research, theoretical dimension of technology adoption and vitality of sentiment analysis, this paper aims to understand the adoption mechanism of Web3 by multi-analytical process of PLS-SEM and Sentiment analysis. For such purpose, causal analysis of technology adoption factors by hypothesis testing was conducted by PLS-SEM and sentiment analysis was conducted through tweets on Web3. By doing so, this study contributes novel hybrid analytical method for behavioral assessment and advances the knowledge towards Web3. This section is followed by theoretical literature of technology adoption factors, then methodology was discussed for the dual analytical scheme. After that, results were presented, and final section elucidated the discussion and conclusion of research on Web3 adoption.

## 2. Literature review
## 2.1. Performance expectancy

The Unified Theory of Acceptance and Use of Technology (UTAUT), a well-known model formulated by [15] is already validated by past researchers in predicting Behavioral Intention to use any information technology. This model illuminated that Performance Expectancy (PE) is one of the influential forecasters of Behavioral Intention. According to [15], PE means the range of individuals' beliefs that their job performance can be improved by using the system. In our research, PE refers to the degree of users' perceptions that Web3 will make them capable of accessing the data from anywhere and controlling their information as well. This means Web3 will help users to retrieve full ownership in controlling their information and having their online privacy.

Following the UTAUT model, a recent study showed the significant effect of PE in foreseeing Behavioral Intention [16]. Surveying 1,562 respondents, the cross-sectional study proved that individuals' actual adoption behavior would significantly turn from users' expected performance while using mobile learning [17]. By collecting 467 responses from the users of digital payment systems in Thailand, the study revealed the impact of PE on users' Behavioral Intention [18]. Moreover, the significant effect of PE as a strong predictor of behavior to use online technology has been found in the prior literature [19]. This view leads to proposing the following hypothesis which can be specified as

**H1:** Performance Expectancy (PE) has a positive influence on Behavioral Intention (INT) to use Web3.

## 2.2. Electronic word-of-mouth

Users rely on fellow users' recommendations more than the content advertising through which they would be more likely to use any technology. Broadly, online reviews from people allow individuals to understand the usefulness of any system that may help them to be aware of systems' usage in the online platform. Thus, Electronic Word-of-Mouth (eWOM) has become an integral part of accelerating information delivery for end-users in the digital business landscape [13], [20]. eWOM is the extent to which former users opine via online their responses either positively or negatively about any technology or service, that would be a reliable source of knowing their best experiences about using such type of technology or service [21].

Using a sample of 314 respondents from Taiwan, a group of researchers proved that positive eWOM profoundly influences consumers' intention in purchasing social networking sites [14]. A study [20] explored that eWOM increases consumers' repurchase intention. A meta-analysis on the effect of eWOM on buying intention identified that the volume of eWOM impacts consumers' buying intention [22]. Based on the 512 effective data, results from an empirical study confirmed eWOM had strong predictive power in explaining consumers' purchase intention [13]. Therefore, it can be hypothesized that:

**H2:** Electronic Word-of-Mouth (eWOM) positively influences Behavioral Intention (INT) to use Web3.

## 2.3. Digital dexterity

Digital Dexterity (DD) refers to the individuals' willingness and abilities to adapt emerging technologies in attaining success in the digital environment. In general, DD means one's broad skills to learn, work, and live in a digital world. A research review by [23] portrayed a DD funnel including three capabilities: Personal Innovativeness, Self and Technology Efficacy. Based on this, we portray DD as the degree of Personal Innovativeness and Technology Self-Efficacy in using Web3. Here, Personal Innovativeness is defined as users' readiness to experiment with Web3 [24], while Technology Self-Efficacy is the extent to which users perceive that they have enough abilities towards Web3 usage [25].

Towards the E-learning context, [10] proved that users' innovative behavior in information technology significantly influences their intention to adopt E-learning. To predict digital competence behavior, a recent study confirmed that Personal Innovativeness is the most dominant predictor of Behavioral Intention [26]. Furthermore, Personal Innovativeness has been found as an influential acceptance determinant of Behavioral Intention in using new technology [27]. On the other hand, results of the 472 analyzed data indicated that users' confidence levels towards using technology strongly impacted their continuance intention towards online learning [28]. Another study showed that Technology Self-Efficacy is a potential predictive factor of technology-based self-directed learning [11]. As digital transformation starts with advanced technology infrastructures and its success depends on human skills in determining the breakeven point towards rapid technological changes and skills demand, DD can be a triggering factor in such a scenario [23]. Based on these findings, we propose another hypothesis as follows:

**H3:** Digital Dexterity (DD) has a positive influence on Behavioral Intention (INT) to use Web3.

## 2.4. Research framework

Based on the analysed literature, the proposed research framework is illustrated in Figure 1, which would be the pathway towards inferences of the study. This research framework integrated three independent variables such as Performance Expectancy, Electronic Word-of-Mouth, and Digital Dexterity. Explicitly, this framework incorporated these three variables as predictive variables of users' Behavioural Intention towards adoption of Web3. The model was validated by a quantitative research survey through PLS-SEM analysis.

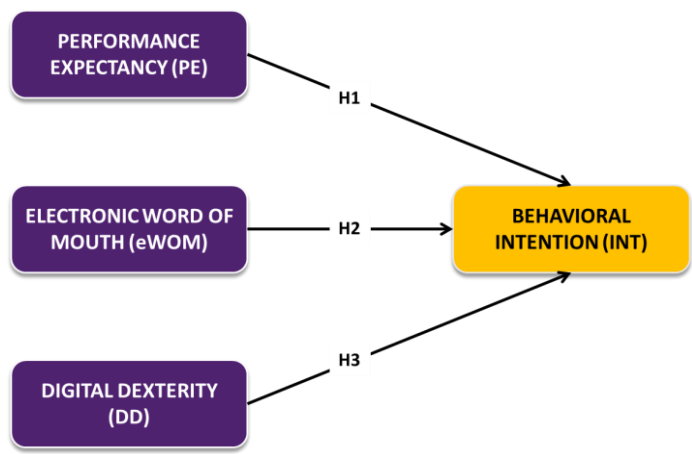

**Figure 1**: Theoretical Framework

## 3. Methodology

By following cross-sectional research designs, we investigated causal relationships among variables. To reach target respondents, we conducted an online survey by targeting the Internet users on Twitter and Facebook. Only those respondents were allowed to fill the survey form who had basic knowledge about Web3 as a future technology tool. Total 167 responses were collected through online questionnaire. We used the snowball sampling technique to reach appropriate and accurate respondents. The survey consisted of two parts: respondents' profiles, and question statements of model variables or items. These questions statements or items were measured through Likert-scale. In our study, a 5-point Likert scale was used to collect the responses against each item or statement of the questionnaire. This scale was scored as 1 = "Strongly Disagree", 2 = "Disagree", 3 = "Neutral", 4 = "Agree", and 5 = "Strongly Agree". Question statements were adopted from previous validated studies. Four instruments to measure Performance Expectancy were adopted from [15]. Electronic Word-of-Mouth was measured using five items from the related study [29]. Digital Dexterity was measured by adopting six items from [11] and [24]. The measure of Behavioral Intention contained five items adopted from [15]. The surveyed questionnaire is also demonstrated in Appendix A. Finally, Partial Least Squares Structural Equation Modeling (PLS-SEM) was used to test the proposed hypotheses through SmartPLS v3. In addition, PLS-predict was assessed to evaluate the model's predictive power.

Towards sentiment analysis, RapidMiner v9.10 was utilized. RapidMiner is a data mining software and widely used for sentiment analysis purpose [8]. Initially data crawling process was generated from "Search Twitter" operator by Twitter API of researcher's Twitter account. To find the tweets of keyword "Web3", only English language tweets were targeted on February 28, 2022. Total 7,410 tweets were crawled during the search query. All collected tweets were dated and generated for same target day i.e., February 28, 2022. Data preprocessing and removing duplicates were contemplated at initial stage [30]. It included filtering the special characters, URLs, and stop words. Afterwards, "Extract Sentiment" operator was used through VADER sentiment tool. Valence Aware Dictionary and sEntiment Reasoner (VADER) is a lexicon and rule-based sentiment analysis tool that is specifically attuned to the sentiments expressed in social media [31], [32]. It takes into consideration the word order and degree modifiers [30]. Then attributes generation operator was edited with the sentiment score to understand the polarity of tweets as if sentiment score > 0 then classify as "positive", if it is less than 0 then term as "negative" and if score = 0, then enlist the tweet as "neutral". Finally, write excel function was used to fetch all the data of sentiment classification.

## 4. Results
## 4.1. Respondents' profiles

According to collected responses, 77% were male and 23% females provided the viewpoint. Among them, 52% had bachelor's degree, 30% had master's and higher education qualification while remaining were below the bachelor's degree. As per the age group, 18-25 counted 22%, 26-34 was 39%, 35-44 was 31% and above 45 was 8%. From respondents' nature of work, 64% had job, 22% dealing their business and remaining were students. Data responses were collected from Twitter and Facebook, mostly users belonged to Southeast Asian countries.

## 4.2. Construct reliability and validity

To evaluate the construct reliability and validity, various analyses were assessed. For example, construct reliability was tested using Cronbach's alpha (α), whereby construct reliability was found to be acceptable with the rule of thumb α > 0.7 [33], as shown in Table 1. We also examined the Composite Reliability (CR) as an estimate of construct reliability. CR value for each construct exceeded the threshold of 0.7 [33]. This result ensured the level of internal consistency for all constructs in our study (Table 1). On the other hand, convergent validity was statistically measured using Average Variance Extracted (AVE). Table 1 presents the convergent validity was confirmed with the AVE value of higher than 0.50 [33]. To test the discriminant validity, we computed the Hetrotrait-Monotrait (HTMT) ratio of correlations criterion. The resulting data with the HTMT ratio less than 0.85 confirmed that discriminant validity of the measurement model was established among reflective constructs [34], as shown in Table 1.

## 4.3. Outer loadings

We conducted outer loadings to determine the reliability of all indicators. The findings of outer loadings (OL) value showed adequate indicator reliability for all constructs as the values of most of the indicators surpassed 0.70 [35]. Table 1 indicates that only PE-1 and DD-1 could not fulfil the criteria, whereas the value of outer loadings ranging between 0.50-0.60 also suggests an acceptable level of indicator reliability [36]. Therefore, PE-1 and DD-1 were reliably accepted for this study.

Additionally, the multicollinearity of the measurements was assessed using the inner Variance Inflation Factor (VIF) for reflective research constructs. The resulting data confirmed that there was no issue of multicollinearity because of having VIF value below 5 for each indicator [33], as presented in Table 1.

**Table 1**
Reliability, Validity, Outer loadings and VIF Results

| Factors | Items | OL | VIF | α | CR | AVE | Discriminant Validity (HTMT) | | | |
|---|---|---|---|---|---|---|---|---|---|---|
| | | | | | | | PE | eWOM | DD | INT |
| PE | PE1 | 0.67 | 1.36 | 0.87 | 0.91 | 0.72 | | | | |
| | PE2 | 0.90 | 3.20 | | | | | | | |
| | PE3 | 0.87 | 2.73 | | | | | | | |
| | PE4 | 0.93 | 4.05 | | | | | | | |
| eWOM | eWOM1 | 0.73 | 1.66 | 0.87 | 0.91 | 0.66 | 0.33 | | | |
| | eWOM2 | 0.82 | 2.25 | | | | | | | |
| | eWOM3 | 0.88 | 2.82 | | | | | | | |
| | eWOM4 | 0.80 | 1.95 | | | | | | | |
| | eWOM5 | 0.81 | 1.95 | | | | | | | |
| DD | DD1 | 0.61 | 1.39 | 0.90 | 0.92 | 0.67 | 0.65 | 0.40 | | |
| | DD2 | 0.92 | 4.18 | | | | | | | |
| | DD3 | 0.84 | 2.60 | | | | | | | |
| | DD4 | 0.76 | 1.92 | | | | | | | |
| | DD5 | 0.92 | 4.47 | | | | | | | |
| | DD6 | 0.82 | 2.27 | | | | | | | |
| INT | INT1 | 0.75 | 1.55 | 0.83 | 0.88 | 0.60 | 0.61 | 0.52 | 0.80 | |
| | INT2 | 0.78 | 1.73 | | | | | | | |
| | INT3 | 0.74 | 1.60 | | | | | | | |
| | INT4 | 0.82 | 1.94 | | | | | | | |
| | INT5 | 0.77 | 1.64 | | | | | | | |

## 4.4. Structural model analysis

The PLS-SEM technique was used to estimate path models and their significance levels to evaluate each hypothesis. The hypotheses were measured on three tools i.e., Beta, T-statistics, and p-value. Standardized regression coefficient (beta or β) which indicates direct effect of an independent variable (here in our study these are PE, eWOM and DD) on a dependent variable (here in our work it is INT) in the path model. Its values range between -1 to +1. Higher value of Beta shows more positive impact of independent variable on dependent variable. T-statistics or "t" is measure of hypothesis testing where t value greater that 1.96 (t > 1.96) is considered as hypothesis acceptance benchmark. Similarly, p-value or "p" is significance value in hypothesis testing where it's value should be less than 0.05 (p < 0.05) to prove the hypothesis acceptance status. In PLS-SEM analysis as depicted in Figure 2, the β values of each hypothesis and R-squared value of model are manifested. The results indicated that the path between Performance Expectancy and Behavioral Intention i.e., Hypothesis 1 (H1) was insignificant with β = 0.151, t-statistics = 1.782, p-value = 0.075. Thus, H1 (PE→INT) was not confirmed and hence rejected for this study. Moreover, users who believed in Electronic Word-of-Mouth towards Web3 were more likely to adopt Web3 in future as results indicated that β = 0.219, ensuring a statistical relationship existed between Electronic Word-of-Mouth and Behavioral Intention with t-statistics = 4.101 and p-value = 0.000. Therefore, H2 (eWOM→INT) was confirmed and accepted for this study. In addition, a statistically significant link between DD and the Behavioral Intention was also found, confirming that Digital Dexterity was an influential precursor element of Web3 adoption behavior. Results for this hypothesis indicated as β = 0.534, t-statistics = 6.552 and p-value = 0.000. This finding sturdily supported H3 (DD→INT). Results of hypothesis testing are displayed in Table 3

Overall, Table 2 and Figure 2 indicate that 54.9% of the variance (R-square value = 0.549) in Behavioral Intention (INT) was occurred due to PE, eWOM, and DD. On the other hand, F-squared or "$f^2$" value clarifies per exogenous variable's effect size in the models. F-Square is the variation in R-Square when an exogenous variable is eliminated from the model. The effect size is measured as if f-square value >=0.02 it is small; when f-square value >= 0.15 is medium and if f-square value >= 0.35 it is large [37]. The findings showed that Digital Dexterity had a large effect size on Behavioral Intention ($f^2 = 0.396$), while small effect size of PE ($f^2 = 0.033$) and eWOM ($f^2 = 0.092$) were found, as illustrated in Table 2.

Table 4 shows results of PLS-Predict analysis where Root Mean Square Error (RMSE) and Mean Absolute Error (MAE) values in the PLS section are lower than Multiple Linear Regression (ML) sections whilst Q square root (Q2) values are greater than ML's respective values, which indicates quite a higher predictive power of our proposed model with non-overfitting problems [38]. PLS-predict results confirmed the predictive validity that resulted as validated prediction of PLS-SEM model.

**Table 2**
F-Square and R-Squared values

| F-square | | | Overall Impact on INT | |
|---|---|---|---|---|
| DD | P.E | eWOM | R Square | R Square Adjusted |
| 0.396 | 0.033 | 0.092 | 0.549 | 0.54 |

**Table 3**
Hypothesis Testing

| Hypotheses | | Path Coefficient (β) | Standard Deviation | T Statistics | P Values | Results |
|---|---|---|---|---|---|---|
| H1 | PE→INT | 0.151 | 0.085 | 1.782 | 0.075 | Rejected |
| H2 | eWOM→INT | 0.219 | 0.053 | 4.101 | 0.000 | Accepted |
| H3 | DD→INT | 0.534 | 0.081 | 6.552 | 0.000 | Accepted |

**Table 4**
Predictive Reliability (PLS-Predict)

|  | PLS | | | LM | | | PLS-LM | | |
|---|---|---|---|---|---|---|---|---|---|
|  | RMSE | MAE | Q² predict | RMSE | MAE | Q² predict | RMSE | MAE | Q² predict |
| INT1 | 0.836 | 0.54 | 0.33 | 0.863 | 0.566 | 0.285 | -0.027 | -0.026 | 0.045 |
| INT2 | 0.851 | 0.557 | 0.306 | 0.901 | 0.6 | 0.222 | -0.05 | -0.043 | 0.084 |
| INT3 | 0.887 | 0.584 | 0.245 | 0.926 | 0.633 | 0.175 | -0.039 | -0.049 | 0.07 |
| INT4 | 0.836 | 0.557 | 0.328 | 0.874 | 0.583 | 0.265 | -0.038 | -0.026 | 0.063 |
| INT5 | 0.819 | 0.57 | 0.321 | 0.835 | 0.606 | 0.295 | -0.016 | -0.036 | 0.026 |

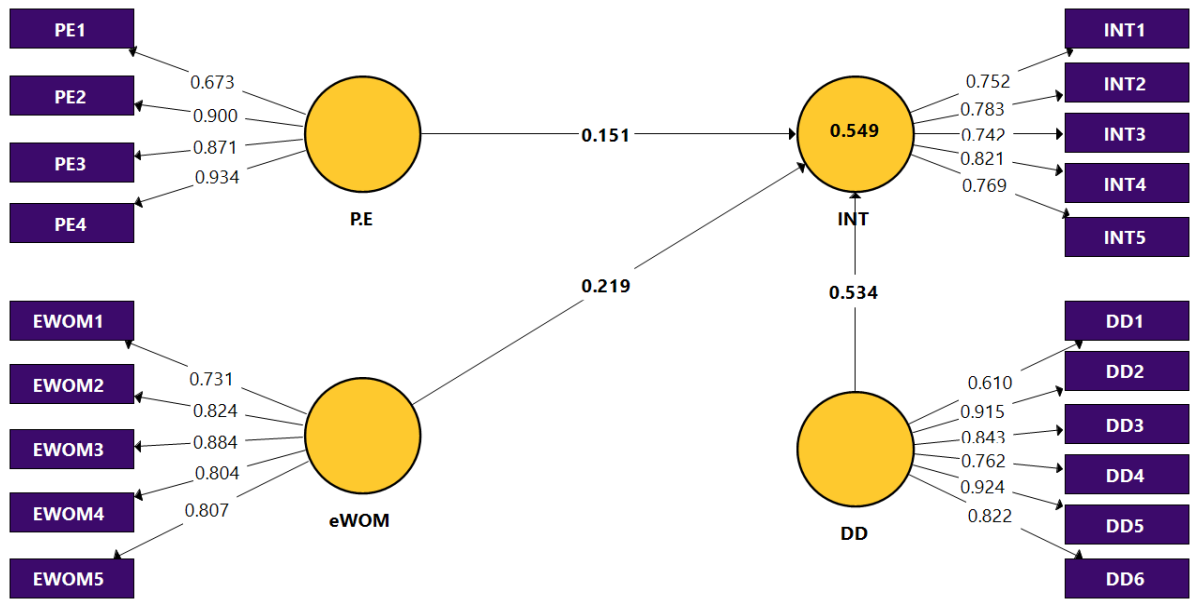

**Figure 2:** PLS-SEM Framework

## 4.5. Sentiment analysis

According to VADER tool, total 3,989 tweets were filtered out with sentiment results. The polarity of tweets was resulted as 2,254 tweets (56.28%) positive, 1,465 tweets (36.73%) neutral and 279 tweets (6.99%) negative. The polarity was measured on behalf of word strings or keywords in tweets that depicted the positive or negative emotions such as "amazing, like, wow, happy, enjoy" etc. for positive emotions and "despicable, sad, worries, lose, drop, fearing" etc. as negative while where no such words were identified by VADER tool, the tweet was termed as neutral. For instance, in Table 5, three tweets and respective sentiment from the analysis are given. To understand the sentiment words used in Web3 tweets, word-clouds are illustrated below. As Figure 3, shows the positive words collection and Figure 4 shows the negative words enlisted in the tweets.

**Table 5**
Sample Tweets with Sentiment Result

| Tweet | Sentiment |
|---|---|
| "Web3 is the first essential step towards a post-scarcity world. One where humanity will be free of strife and conflict over resources." | Positive |
| ""Decentralisation" they said. You go from a government spying on you and doing whatever they want to web3 companies freezing you based on where you're from. Despicable" | Negative |
| "We joined other Web3 pioneers earlier this month to speak at axtech, one of the first NFT events in Sweden" | Neutral |

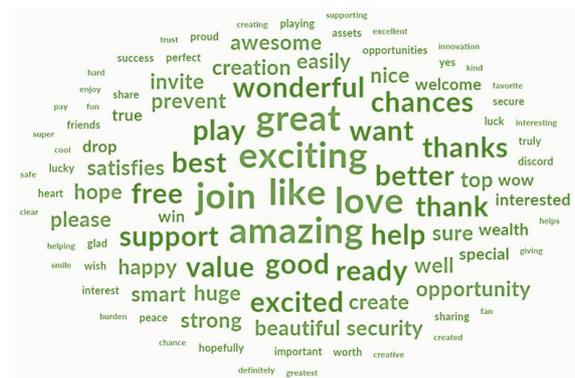

**Figure 3:** Positive Word Collection

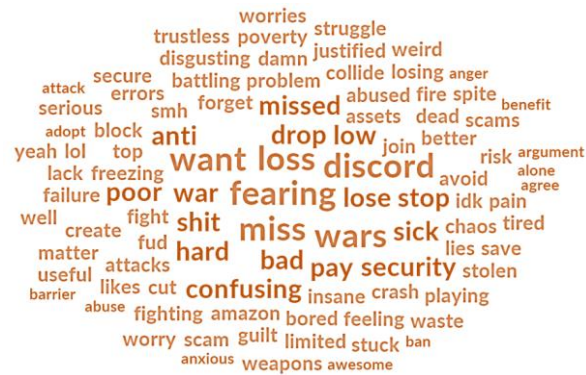

**Figure 4:** Negative Word Collection

## 5. Conclusion

The purpose of this study was to understand Web3 adoption behavior and to be aware of general sentiment inclination. For such sense, PLS-SEM and Twitter sentiment analysis were conducted. The results of both analysis techniques explored the different perspectives of Web3 adoption. In PLS-SEM causal behavioral analysis, the hypothesized model revealed the impact on behavior to adopt the Web3 is 54.9% with the prominent effect of Digital Dexterity. Similarly, users who consider the Electronic Word-of-Mouth as instigating tool to use web services will also be likely to adopt the

Web3. While surveyed users perceived that Performance Expectancy of Web3 at this stage will not be privileged for enhancing their performance at work or personal life therefore Performance Expectancy evolved as non-significant element in the research framework. On other hand, Twitter sentiment analysis presented the polarity of the majority of tweets as positive. While around 7% tweets reiterated the negative sentiment towards Web3.

Our results are validating the hybrid analysis of PLS-SEM with sentiment assessment to comprehend the broader spectrum of studying phenomenon. Understanding the behavior at pre-adoption stage, sentiment analysis could be suitable evaluation pattern with PLS-SEM studies. Hypothesis testing in our study validated the relationships of Digital Dexterity and Electronic Word-of-Mouth towards Behavioral Intention as practiced in previous studies [13], [14], [10], [11], [18], [39] while Performance Expectancy being non-significant factor towards Behavioral Intention has differed the inferences from literature [10], [16]. Twitter Sentiment analysis techniques and results were also validated as per the previous researches in this domain [8], [9], [30], [31].

The inferences explored that Web3 is distinctive and future-oriented technology however at this moment it is merely being used for marketing or promoting businesses whereas actual benefits are not yet handy. For this reason, its polarity among online folks is positive but future usability at workplace or in personal life affairs is not comprehensible and hence considered as ineffective at present. Many of analyzed tweets for sentiment polarity were also depicting the promotion and marketing impression for NFTs and cryptocurrency tools. While most of the futuristic benefits of Web3 are mainly described on social media feeds and tweets, and less practicality in real life is demonstrated. Therefore, users are, somehow, unwilling towards Web3 integration into Internet. While it also shows that Internet users with Digital Dexterity i.e., elevated level of innovativeness and technology awareness will highly be likely to become early adopters of Web3.

The study is the pioneer contribution to the Web3 literature in terms of sentiment analysis and behavioral assessment. The sentiment analysis inferences will support the stakeholders to make the wise decisions regarding the Web3 implementation. It showed the positive buzz among internet users towards Web3 but still its usefulness as per surveyed results, has not been perceived by the users. To spread more words about Web3 performance and feasible advantages can be the "lesson learnt" from our sentiment and causal analysis results. Through hypothesized framework results, this research has revealed that Web3 adoption will relate to better understanding of digital competence and technological prowess, whereas Web 2.0 had not required such complexity of technological awareness towards its functionality. Meanwhile positive buzz will also attract multiple factions of society to join the Web3 band through various interest groups such as NFTs, Crypto, data privacy initiatives etc. However, it is also orchestrated that Web3 is considered as "hype" over Internet to attract potential stakeholder. It also includes the programmers and tech-gurus, who are preparing and training the decentralized technologies tools and techniques for better future.

Personalization of Web3 has already been integrated through numerous ways such as connectivity of IoT devices in smart home, user-data tracking and interaction of websites, edge computing and semantic web. When Web 2.0 was came to limelight in 2000-2010 era, the concept of semantic web was also existed at that time with the expression of Semantic Edge [40]. Semantic as a word refers to Meaning or Logic of respective phenomenon and Semantic Web could overtly direct the phenomenon of deriving the meaning of Web activities through users' data and interactions. It can be described that the mechanism of Semantic Web is the backbone of term "Web3" [40]–[42]. Similarly, another expression in Web3 is decentralization which entails the Decentralized Finance or "DeFi" (i.e., open banking system based on Distributed Ledger Technology or "DLT") [43] and Cryptocurrencies/Bitcoins in the society, would require the regulations, time and digital infrastructure to be implemented. However, besides all such developments, the digital inequality will be increased with the time across the globe. Developing and underdeveloped regions from global south needs the digital infrastructure (in shape of Web 2.0) to infuse into education, health, transport, and communication system for sustainable development. It would be appropriate to fully integrate the Web 2.0 prior to sailing on the Web3 ocean.

It is the initial research work on behavioral modelling of Web3 and delivers the resourceful viewpoint for forthcoming research. Exploring several limitations of this study may produce noteworthy references for further study. Firstly, we compiled a small set of data based on the snowball sampling, which could not provide a broad measure of respondents' Behavioral Intention.

Therefore, a larger sample size used in future studies could draw better inferences [17]. Moreover, we only emphasized three predictive factors (i.e., PE, eWOM, and DD) of users' Behavioral Intention, while focusing on different factors (i.e., perceived authenticity, perceived value etc.) that would explain more insight about the Behavioral Intention to adopt Web3. Regarding sentiment analysis, we have practiced the VADER assessment tool, while Naïve Bayes sentiment model could be implemented to train and test the tweets data for sentiment analysis. Further researchers may evaluate such factors and compare the variances with the mechanisms of our study. Also, they can compare our model with different geographical contexts for better generalizability of the current findings.

# Appendix A. Questionnaire

## Variables' Questionnaire Items

**Performance Expectancy (PE)**

(PE1) - I would find Web3 useful in my task.

(PE2) - Using Web3 will enable me to accomplish tasks more quickly.

(PE3) - Using Web3 will increase my productivity.

(PE4) - If I use Web3, I will increase my chances of getting a raise.

**Electronic Word-of-Mouth (eWOM)**

(eWOM1) - People's recommendations on the internet regarding Web3 are useful for me.

(eWOM2) - People's recommendations on the internet about Web3 influence me to use it.

(eWOM3) - People's recommendations on the internet about Web3 would increase my interest in finding out more.

(eWOM4) - I will decide to use Web3 based on peoples' recommendations I receive.

(eWOM5) - The data about Web3 on the internet meets my information needs.

**Digital Dexterity (DD)**

(DD1) - I know how to use Web3 on my own.

(DD2) - I believe I have enough knowledge of using Web3.

(DD3) - I would look for ways to experiment with Web3.

(DD4) - I want to experiment with Web3.

(DD5) - I am not hesitant to try out Web3.

(DD6) - I am usually at early step to try out new information technology like Web3.

**Behavioral Intention (INT)**

(INT1) - Assuming I can access the Web3 system, I intend to use it.

(INT2) - Given that I have access to the s Web3 system, I predict that I would use it.

(INT3) - I intend to use the Web3 system in the next months.

(INT4) - I predict I would use the Web3 system in the next months.

(INT5) - I plan to use the Web3 system in the next months.